\documentclass[aps, prd, twocolumn, superscriptaddress, nofootinbib]{revtex4}
\usepackage{graphicx}% Include figure files
\usepackage{dcolumn}% Align table columns on decimal point
\usepackage{amssymb}% outlined letters
\usepackage{amsmath,amssymb,amsfonts}
\usepackage[normalem]{ulem}
\usepackage{tabularx,ragged2e,booktabs,caption}

\begin{document}

%Macros
\newcommand{\Eq}[1]{\mbox{Eq. (\ref{eqn:#1})}}
\newcommand{\Fig}[1]{\mbox{Fig. \ref{fig:#1}}}
\newcommand{\Sec}[1]{\mbox{Sec. \ref{sec:#1}}}

\newcommand{\PHI}{\phi}
\newcommand{\PhiN}{\Phi^{\mathrm{N}}}
\newcommand{\vect}[1]{\mathbf{#1}}
\newcommand{\Del}{\nabla}
\newcommand{\unit}[1]{\;\mathrm{#1}}
\newcommand{\x}{\vect{x}}
\newcommand{\ScS}{\scriptstyle}
\newcommand{\ScScS}{\scriptscriptstyle}
\newcommand{\xplus}[1]{\vect{x}\!\ScScS{+}\!\ScS\vect{#1}}
\newcommand{\xminus}[1]{\vect{x}\!\ScScS{-}\!\ScS\vect{#1}}
\newcommand{\diff}{\mathrm{d}}

\newcommand{\be}{\begin{equation}}
\newcommand{\ee}{\end{equation}}
\newcommand{\bea}{\begin{eqnarray}}
\newcommand{\eea}{\end{eqnarray}}
\newcommand{\vu}{{\mathbf u}}
\newcommand{\ve}{{\mathbf e}}

%=====================================================================
%=====================================================================
%=====================================================================

\title{Planck-scale dimensional reduction without a preferred frame}

\newcommand{\addressImperial}{Theoretical Physics, Blackett Laboratory, Imperial College, London, SW7 2BZ, United Kingdom}
\newcommand{\addressRoma}{Dipartimento di Fisica, Universit\`a “La Sapienza”
and Sez. Roma1 INFN, P.le A. Moro 2, 00185 Roma, Italia}

\author{Giovanni Amelino-Camelia}
\affiliation{\addressRoma}
\author{Michele Arzano}
\affiliation{\addressRoma}
\author{Giulia Gubitosi}
\affiliation{\addressRoma}
\author{Jo\~{a}o Magueijo}
%\email{magueijo@ic.ac.uk}
\affiliation{\addressImperial}
%\affiliation{\addressRoma}

\date{\today}

\begin{abstract}
Several approaches to quantum gravity suggest that the standard description of spacetime as probed at low-energy, with four dimensions, is replaced in the Planckian regime by a spacetime with a spectral dimension of two. The implications for relativistic symmetries can be momentous, and indeed the most tangible picture for ``running'' of the spectral dimension, found within Horava-Lifschitz gravity, requires the breakdown of relativity of inertial frames. In this Letter we incorporate running spectral dimensions in a scenario that does not require the emergence of a preferred frame. We consider the best studied mechanism for deforming relativistic symmetries whilst preserving the relativity of inertial frames, based on a momentum space with curvature at the Planck scale. We show explicitly how running of the spectral dimension can be derived from these models.
 \end{abstract}

\keywords{}
\pacs{}

\maketitle

%=====================================================================
%=====================================================================
%=====================================================================

%\section{Introduction}
{\bf 1. Introduction.}
One of the most robust predictions of quantum-gravity is that spacetime itself should acquire quantum properties, when probed at the Planck scale ($\simeq 10^{28}eV$). Over the past decade it was gradually appreciated that two key issues deserve priority in investigations of quantum models of spacetime:
\begin{itemize}
\item
What is the fate of relativistic symmetries in the Planck-scale description of spacetime?
\item
Is spacetime still four-dimensional in the Planck-scale regime?
\end{itemize}
The second question may appear to be ill-defined, since our intuitive notion of spacetime dimensionality is based on properties of purely classical geometries. The most intuitive such notion is the Hausdorff dimension, captured by the scaling exponent of the volume of a sphere, but a smooth classical geometry is a
prerequisite for considering spheres and an {\it uncontroversial} concept of their volume. This has led quantum-gravity researchers to employ an alternative definition: the spectral dimension. This concept is encoded in the spectral properties of the scalar Laplacian for the theory of interest. For smooth classical spacetimes the spectral dimension coincides with the Hausdorff dimension. In a quantum geometry the latter is in general inapplicable, but the spectral dimension of spacetime is still well-defined.

Interestingly, as  the spectral dimension criterion became adopted in a growing number of approaches, it emerged that rather generically the spectral dimension in the UV\ regime is smaller than 4 (e.g.\cite{ASTRID} and references therein). It is particularly intriguing that some of the most studied, but ostensibly very different, quantum gravity theories predict that the value of the spectral dimension in the UV\ is 2. This conclusion finds support in the CDT (Causal-Dynamical-Triangulation) approach \cite{Ambjorn:2005db}, Asymptotic Safety \cite{Litim:2003vp}, Horava-Lifshitz (HL) gravity \cite{Horava:2009if}, and Loop Quantum Gravity (LQG) \cite{modesto}.

Irrespective of the alleged UV dimensional reduction phenomenon, the fate of relativistic symmetries in the Planckian regime has attracted interest from other angles (see e.g.\cite{Mattingly:2005re, AmelinoCamelia:2008qg,AmelinoCamelia:2009pg}). Relativistic symmetries may be left unscathed by the new structures at the Planck scale (e.g.\cite{causalsets}), but there are at least two other possibilities. Planck-scale effects may {\it break} relativistic invariance, introducing a preferred-frame \cite{grbgac,gampul,urrutia,Moffat:1992ud,Albrecht:1998ir}; or they may  {\it deform}  the relativistic symmetry transformations, preserving the relativity of inertial frames \cite{dsruno,dsrdue,jurekDSRfirst,leejoaoPRLdsr,leejoaoPRDdsr,jurekDSMOMENTUM}.
In this Letter we contribute to the understanding of the interplay between spectral dimensional reduction and the fate of relativistic symmetries at the Planck scale.

It is evident that any model of spacetime with dimensional reduction must bring relativistic transformations under scrutiny \cite{carlip,prlSILKE}. Yet, in most~studies the analysis is confined to the perspective of a single observer, without mention of how a boosted observer would describe the same phenomenon. An exception is found in HL gravity, where an explicit breakdown of the equivalence of inertial observers is vividly manifest \cite{Horava:2009if,prlSILKE}. The fate of relativistic invariance in CDT, Asymptotic Safety and LQG remains the subject of a lively debate (e.g.\cite{urrutia,Calmet:2010tx,Jordan:2013awa}). We hope to contribute to this debate by showing that the phenomenon of running spectral dimension arises naturally within the most studied mechanism for deformation of relativistic symmetries, preserving the relativity of inertial frames. The mechanism assumes that momentum space is curved at the Planck scale, and remarkably it can easily describe the topical case of a two-dimensional UV regime. In our closing remarks we discuss the significance of these findings.

{\bf 2. Preferred-frame scenarios for running spectral dimension.}
We start by summarizing a few facts about preferred-frame scenarios. The results collected here are all essentially known (\cite{Sotiriou:2011aa} and references therein), but some have not been previously spelled out as explicitly as we shall do. They will be useful to contrast preferred-frame and frame-invariant models, the latter being our main focus of interest.

A useful starting point is a modified-Laplacian with Euclideanized (``Wick-rotated'')  form:
\begin{equation}
D_{E}=-\partial_{t}^{2}-\nabla^{2}
-\ell_t^{2\gamma_{t}}\partial_{t}^{2(1+\gamma_{t})}-\ell_x^{2\gamma_{x}}\nabla^{2(1+\gamma_{x})}
\label{joc0}
\end{equation}
where $\gamma_t$ and $\gamma_x$ are
dimensionless parameters, and $\ell_t$ and $\ell_x$ are parameters with dimensions of length (usually assumed to be of the order of the inverse of the Planck-scale).
Modified Laplacians with Euclidean form given by (\ref{joc0}) are relevant in scenarios where Planck-scale effects break relativistic invariance, often labelled LIV (Lorentz Invariance Violating). The transition from IR to UV in the chosen Laplacians is not relevant for this paper, only the asymptotic forms.

The spectral dimension is the effective dimension probed by a fictitious diffusion process governed by the ``Wick rotated'' Laplacian operator of the theory.
The core features are encoded in the average return probability, given by
\begin{equation}
P(s) \propto \int dE \,dp \, p^{D-1} \, e^{-s\Omega(E,p)}\, ,\label{jocreturn}
\end{equation}
where $s$ is a fictitious ``diffusion time", $p$ is the modulus of the spatial momentum, $D$ is the number of spatial dimensions in the IR regime, for which we shall often assume $D=3$,  and $\Omega(p)$ is the momentum-space representation of the Laplacian operator, which for (\ref{joc0}) is:
\begin{equation}\Omega_{\gamma_t \gamma_x}(E,p)= E^{2}+p^{2}+\ell_{t}^{2\gamma_{t}}E^{2(1+\gamma_{t})}+\ell_{x}^{2\gamma_{x}}p^{2(1+\gamma_{x})}\, .\label{eq:Omegagammatgammax}
\end{equation}
In (\ref{jocreturn}) we disregarded an overall numerical coefficient which does not affect the spectral dimension analysis. The spectral dimension in the UV regime, $d_{S}(0)$, is obtained from the return probability by computing
\begin{equation}
d_{S}(0)= -2 \lim_{s\rightarrow 0}\frac{d\ln P(s)}{d\ln(s)}\, ,
\label{joc7}\end{equation}
and the spectral dimension ``runs'' whenever $d_{S}(0) \neq D+1$. The spectral dimension in the IR regime, $d_{S}(\infty)$, is obtained from a similar formula but with limit $s\rightarrow\infty$, and in all theories considered here $d_{S}(\infty)=D+1$.

To compute $d_{S}(0)$ for the LIV models characterized by Eq.(\ref{eq:Omegagammatgammax}) we can use \cite{Horava:2009if,Sotiriou:2011aa}
\begin{eqnarray}
&& f(z) \propto \int dx x^{n} \, e^{-z (x^2+\alpha x^{2\beta})}
\label{jocformula}\\
&&\,\,\,\,\,\,\,\,\,\,\,\,\,\,\,\,\,\,\,\,\,\,\,\,\,\,\,\, \,\,\,\,\,\,\,\,\,\,\,
\,\,\,\,\,\,\, \Longrightarrow \,\,\,
 \lim_{z\rightarrow 0}\frac{d\ln f(z)}{d\ln(z)}= -\frac{n+1}{2\,\beta}
\nonumber
\end{eqnarray}
so that:
\begin{equation}
d_{S}(0)=\frac{1}{1+\gamma_t}+\frac{D}{1+\gamma_x}%~\frac{bla}{bla} \gamma_t \gamma_x D
\, .
\label{joc1}
\end{equation}
The LIV model with $\gamma_t =0$ and $\gamma_x=2$ describes
HL gravity \cite{Horava:2009if,prlSILKE}, and indeed gives $d_{S}(0)=2$ for $D=3$.  Eq.(\ref{joc1}) generalizes this result.

In Ref.\cite{measurematters} we established a
correspondence between the UV spectral dimension
of spacetime and the UV Hausdorff dimension of momentum space for LIV models
with $\gamma_t =0$ and general $\gamma_x$.
We now prove that the argument applies for arbitrary values of
$\gamma_t $ and $\gamma_x$. For this purpose we adopt a change
of integration variables which in the UV takes the form:
\bea
\tilde E&\propto&E^{1+\gamma_{t}} \nonumber\\
\tilde p&\propto &p^{1+\gamma_{x}}
\eea
whereas in IR it leaves momentum space unchanged (the transition between the two regimes is irrelevant for the argument).
Then {\it any} integral over momentum space involving a
function of the UV-modified Laplacian will be converted
into an integral
of the same function of the unmodified Laplacian, but with a suitably
UV-modified integration measure.
For example, Eq.(\ref{jocreturn}) becomes:
\begin{equation}
P(s)  \propto \int {d \tilde E \,d\tilde p\, \,\tilde p^{\frac{D-\gamma_{x}-1}{\gamma_{x}+1}}\,\tilde  E^{-\frac{\gamma_{t}}{1+\gamma_{t}}}}\,e^{-s(\tilde E^{2}+\tilde p^{2})}\, ,\label{eq:ReturnProb1}
\end{equation}
up to terms that are negligible in the UV regime. We notice that the relevant integration measure factorizes in $E$ and $p$, leading to a valuable intuitive characterization of (\ref{joc1}).
The energy integration has measure $d\tilde E\,\tilde E^{\frac{1}{1+\gamma_{t}}-1}$ suggesting that the effective UV Hausdorff  dimension of energy space is $1/(1+\gamma_{t})$,
whereas the momentum integration measure is $d \tilde p \, \tilde p^{\frac{D}{1+\gamma_{x}}-1}$ suggesting $D/(1+\gamma_{x})$ Hausdorff  dimensions. These match the two terms in (\ref{joc1}), thereby generalizing the argument in \cite{measurematters}.

{\bf 3. Running spectral dimension without a preferred frame.}
Preferred-frame LIV scenarios, such as the one contained in HL gravity, provide a compelling description of Planck-scale dimensional reduction,
including the topical case of a 2-dimensional UV regime. We now show that an equally encouraging picture of the same phenomenon can be found in scenarios where the relativistic symmetries are deformed, without spoiling the relativity of inertial frames. These are often dubbed ``DSR'' (Doubly, or Deformed, Special Relativity). Remarkably, we need look no further than the simplest such scheme~\cite{dsruno,jurekDSMOMENTUM,principle}, which is based on the assumption that momentum space has de Sitter geometry, with the Planck scale playing the role of its curvature scale.

We start by highlighting the relevant features of the model, referring the interested reader to the copious literature for more detail (e.g.\cite{dsruno,jurekDSMOMENTUM,principle} and Ref.\cite{dsEUCLID} for the Euclideanization prescription). For definiteness let us use a coordinatization such that the (Wick-rotated) de Sitter metric on momentum space is:
$$
ds^2=g^{\mu\nu}dp_\mu dp_\nu
=dE^2+e^{2\ell E}\sum^{D}_{j=1}dp_j^2 \,\, .$$
%$$
%g^{\mu\nu}= \delta^{\mu 0}\delta^{\nu 0}
%+\delta^{\mu \nu}
%(\delta^{\mu 0}\delta^{\nu 0}-1)e^{2\ell E }$$
The fact that these theories do not pick a preferred frame, but do require a deformation of relativistic transformation laws, is a direct  consequence of the fact that de Sitter space is a maximally symmetric geometry. One of several equivalent ways of introducing ordinary special relativity starts from the isometries of a Minkowski momentum space and then derives the transformation laws of spacetime coordinates by consistency \cite{principle}. The isometries of de Sitter momentum space can be seen as a deformation of the isometries of Minkowski momentum space, and as a result a theory built upon the isometries of de Sitter momentum space  is as ``relativistic'' as special relativity, i.e. preserves equally well the relativity of inertial frames. However it will entail deformations of the transformation laws among observers.

Most notably, the ordinary special-relativistic law for the action of boosts on momenta, $[N_j,p_i]=\delta_{ij} E$ and $[N_j,E]=p_{j}$, must be replaced by laws which for our coordinatization of de Sitter momentum space take the form~\cite{dsruno,jurekDSMOMENTUM}
$$[N_j,p_i]=\delta_{ij}\left(\frac{1}{2\ell}(1-e^{-2\ell E})+\frac{\ell}{2}|\vec p|^{2}\right)-\ell p_{j} p_{i}$$
$$[N_j,E]=p_{j}.$$
It is easy to see how this affects the analysis of the spectral dimension of spacetime. Regarding the measure of integration over momentum space, we should include the usual factor $\sqrt{-g}=e^{D\ell E}$. Concerning the choice of the momentum-space representation of the deformed Laplacian operator we must require that it is invariant under relativistic transformations, i.e. under all the isometries of  de Sitter momentum space. The literature~\cite{dsruno,jurekDSMOMENTUM,principle} emphasizes a family of scalars with the desired invariance properties: arbitrary functions $f({\cal C}_\ell)$ of the invariant
$${\cal C}_\ell = \frac{4}{\ell^{2}}\sinh^{2}\left(\frac{\ell E}{2}\right)+e^{\ell E}|\vec p|^{2}.$$
In analogy with our parameterization of LIV\ scenarios with exponents $\gamma_t$ and $\gamma_x$ (cf. Eq.(\ref{eq:Omegagammatgammax})), we adopt the momentum space representation of the Laplacian $\Omega={\cal C}_\ell
+ \ell^{2\gamma} {\cal C}_\ell^{1+\gamma}$, parameterized by a single $\gamma$.

With these ingredients in hand, the return probability $P(s)$ is therefore:
\begin{equation}
P(s) \propto  \int dE dp \, p^{D-1} \, e^{D \ell E} \, e^{-s \left({\cal C}_\ell
+ \ell^{2\gamma} {\cal C}_\ell^{1+\gamma}\right) } \, .
\label{eq:kappaReturnProbability}
\end{equation}
This can be evaluated analytically using the fact that we are interested exclusively in the UV spectral dimension. We can therefore introduce changes of coordinates which
are trivial in the IR, but
simplify adequately our integrals in the UV. Specifically, in the UV we can first introduce $\tilde E=e^{\ell E/2}/\ell$ and $\tilde p=e^{\ell E/2}p$ and then introduce ``polar coordinates'' $\tilde E= r\cos \theta$ and $\tilde p = r\sin \theta$. This allows us to rewrite the return probability (\ref{eq:kappaReturnProbability}) in the UV as
\be
P(s)\propto \int dr\,  r^{2D-1} e^{-sr^{2(\gamma+1)}}\, .
\ee
Using (\ref{jocformula}) we therefore get:
\begin{equation}
d_{S}(0)=\frac{2D}{1+\gamma}
%~\frac{bla}{bla} \gamma_t \gamma_x D
\, .
\label{joc8}
\end{equation}
We have  tested this result numerically by applying the definition
of spectral dimension (\ref{joc7}) directly to (\ref{eq:kappaReturnProbability}). Remarkably, as with the celebrated result for HL theory (giving $d_{S}(0)=2$ for $D=3$, $\gamma_x=2$ and $\gamma_t =0$), our key equation  (\ref{joc8}) for the DSR\ model implies a 2-dimensional UV regime for  $\gamma =2$ and $D=3$. We will discuss this important point in our closing remarks.

It is significant that the identification of the UV spectral dimension of spacetime and the UV Hausdorff dimension of momentum space, which we proved for LIV theories, extends to DSR {\it but with a notable difference}. If after performing the changes of coordinates used above to evaluate $P(s)$ we perform a final replacement $\hat r= r^{1+\gamma}$, then  {\it any} integral over momentum space involving a function of the deformed Laplacian is converted, in the UV, into an integral of the same function of the unmodified Laplacian, but with integration measure:
\be
d\mu \propto \hat r^{\frac{2D}{1+\gamma}-1}\, d\hat r
\ee
where $\hat r^2={\hat E}^2+{\hat p}^2$ now represents the Laplacian in momentum space. Therefore, the UV Hausdorff dimension of momentum space in a picture that leaves the Laplacian unmodified once again exactly matches the spectral dimension of spacetime in the UV. However, the measure now does not factorize into a function of energy and a function of momentum, an important difference with regards to LIV models.

{\bf 4. Significance of our findings.}
The implications of our findings must be assessed against the background of the present status of quantum gravity. Due to a dearth of experimental clues, a variety of approaches are being developed, which not only lead to different predictions but also ``speak'' different languages. It is striking that the issues of dimensional reduction and the fate of relativistic symmetries are among the few topics that cross the boundaries between the various approaches. Their interplay could therefore have a key role in  characterizing similarities and differences among competing theories.

There are a few lessons to be gleaned from our findings. First note that we can collect the results for LIV/preferred-frame models (Eq.(\ref{joc1})) and for DSR/relativistic theories (Eq.(\ref{joc8})) in a general formula:
\be\label{dsgen}
d_S(0)=\frac{{\cal D}_t}{1+\gamma_t}+\frac{{\cal D}_x}{1+\gamma_x}\, .
\ee
 For LIV theories we have  ${\cal D}_x=D$ and ${\cal D}_t=1$ (with $\gamma_t=0$ for HL theory). For DSR we find instead   ${\cal D}_x={\cal D}_t=D$ and $\gamma_x = \gamma_t =\gamma$. We conjecture that all quantum gravity theories fit into Eq.(\ref{dsgen}). Table I shows some quantitative aspects of this formula, pertaining to the ``spectrum'' of UV dimensions obtained by assigning to the exponents $\gamma,\gamma_x,\gamma_t$ small integer values.

It is noteworthy that deformed-relativity with de Sitter momentum space can {\it never} lead to a spectral dimension of 4 in the UV regime. For all non-vanishing integer values of $\gamma$ we are led to dimensional reduction, and for $\gamma=0$ we have $d_{S}(0)=6$, implying UV ``super-diffusion"  \cite{superdiff}. Even more striking is the numerology pertinent to the special case $d_{S}(0)=2$. We note that $d_{S}(0)=2 $  and $d_{S}(0)=1.5$ are the only values of $d_S(0)$ that admit both LIV and DSR descriptions, and that $d_{S}(0)=2 $ is the only case with three different entries in our table. One is HL theory with $\gamma=2$, another is the LIV theory with $\gamma_t=\gamma_x=1$, and finally we have DSR with $\gamma=2$. It is fair to expect that there must be something deep about this case, yet to be fully uncovered. It may be significant that LIV theories with $d_S(0)=2$ are closely linked with the emergence of scale invariant cosmological fluctuations \cite{rsdSKY,dsrflucts}.  We conjecture that this connection is more general than so far appreciated.

\begin{table}[ht]
\begin{center}
\begin{tabular}{  l || c | c | c | c | c | c |   c | c |  c | c  | c  }
$d_{S}(0)$ & \,\! 6 \,\! & \,\! 4 \,\!& 3.5 & \,\! 3 \,\! & 2.5 & \,\,\,\,\,\,\, 2 \,\,\,\,\,\,\,  & 1.75 & \,\! 1.5 \,\! & 4/3  & 1.25\\
\hline
$\,\, \gamma$  & 0 & $\, $  & $\, $  & 1 & $\, $  & 2 &  $\, $ & 3    & $\, $  &
%\hline
%MDSR & 1  & $\gamma$ & $D$ &  $\gamma$\\
\end{tabular}
\end{center}
\end{table}

\vskip -0.8cm

\begin{table}[ht]
\begin{center}
\begin{tabular}{  l || c | c | c | c | c |  c| c | c |  c | c  |c  | c  }
$d_{S}(0)$ & \,\! 6 \,\! & \,\! 4 \,\!& 3.5 & \,\! 3 \,\! & 2.5 & \, 2 \, & \, 2 \, & 1.75 & \,\! 1.5 \,\! & 4/3  & 1.25\\
\hline
$\,\, \gamma_x$ &    & 0  & 0  &    & 1  & 2  &   1&  1 &   2 & 2& 2\\
\hline
$\,\, \gamma_t$ &   & 0 & 1 & & 0 &   0 & 1& 2 &  1 & 2& 3
%\hline
%MDSR & 1  & $\gamma$ & $D$ &  $\gamma$\\
\end{tabular}
\caption{\footnotesize{The values of $d_{S}(0)$ that can be obtained by assigning small integer values to $\gamma$ (DSR case, top table), and to $\gamma_{x}$ and $\gamma_{t}$ (LIV case, bottom table). We assume $D=3$, so that in all the cases the IR value of the spectral dimension is $d_{S}(\infty)=4$. }}
\end{center}
\end{table}

We also conjecture that our results could play a role in giving shape to Born's pioneering vision proposed in 1938 \cite{born}. Born argued that  a quantum theory of gravity could only be successful if one allowed for momentum space to be curved, an expression of ``Born reciprocity''. After being dismissed for decades, Born's idea is undergoing a revival. In the context of our Letter we highlight the following. It is well known that spacetime curvature produces running of the spectral dimension {\it in the infrared} (e.g. \cite{Benedetti:2009ge} and references therein). We now find that momentum space curvature induces running of the spectral dimension {\it in the ultraviolet}, and that this may be focal in describing the phenomenon without introducing preferred frames. A more subtle implementation of Born's vision is thus suggested: it could be that the two types of curvature (that of momentum space and that of spacetime) are only relevant in dual regimes, characterized by the UV and IR limits.

In closing we comment on implications for concrete quantum-gravity approaches. Our results are evidently applicable
to the recently proposed relative-locality approach \cite{principle},
which is centered on curvature of momentum space. Several studies have linked de Sitter momentum space  to the $\kappa$-Minkowski noncommutative spacetime \cite{dsruno,jurekDSMOMENTUM,dsEUCLID,darione,micales}.
It is therefore relevant that the results obtained in
 Ref.\cite{darione} for running of the spectral dimension in
 $\kappa$-Minkowski spacetime are described by the special case $\gamma=1$
of our more general result (\ref{joc8}).
Our study may also contribute to a better understanding of the CDT approach. Early work implied that CDT models contained a preferred-frame, just like HL theory. In particular, a perfect match with HL theory for the full $d_S(s)$ function ({\it i.e.} $-2 d\ln P(s)/d\ln(s)$) was exhibited in \cite{prlSILKE}.
Two recent results have weakened this claim. Firstly, Ref.\cite{ASTRID} showed that, whereas  $d_S(0)$ and $d_S(\infty)$ are unproblematic, the full function $d_{S}(s)$ cannot be used for a reliable comparison of different scenarios. Then, Ref.\cite{Jordan:2013awa} provided direct evidence that the CDT results are independent of the spacetime foliation.  Our description of running spectral dimension without a preferred frame shows that it is indeed not necessary to introduce preferred frames to describe the phenomenon.
%Our model should be seen as the right language in which to couch the CDT phenomenology reported in \cite{Ambjorn:2005db}.

{\it Acknowledgments.}
We thank Macarena Lagos for a critical reading of the manuscript. GAC, MA, and GG were supported in part by the John Templeton Foundation. The work of MA was also supported by the EU Marie Curie Actions through a Career Integration Grant. JM was funded by STFC through a consolidated grant.

\end{document}